 \documentclass[preprint2]{aastex} 

\slugcomment{}

\shorttitle{Correlations between CME parameters and sunspot
activity}
  \shortauthors{Z. L. Du}

\begin{document}
\title{Correlations between CME Parameters and Sunspot Activity}

\author{Z. L. Du\altaffilmark{}}
\affil{Key Laboratory of Solar Activity, National Astronomical
Observatories, Chinese Academy of Sciences, Beijing 100012, China}
\email{zldu@nao.cas.cn}

\begin{abstract}
Smoothed monthly mean coronal mass ejection (CME) parameters
(speed, acceleration, central position angle, angular width, mass
and kinetic energy) for Cycle 23 are cross-analyzed, showing a
high correlation between most of them. The CME acceleration ($a$)
is found to be highly correlated with the reciprocal of its mass
($M$), with a correlation coefficient $r=0.899$. The force ($Ma$)
to drive a CME is found to be well anti-correlated with the
sunspot number ($R_\mathrm{z}$), $r=-0.750$. The relationships
between CME parameters and $R_\mathrm{z}$ can be well described by
an integral response model with a decay time scale of about 11
months. The correlation coefficients of CME parameters with the
reconstructed series based on this model
($\overline{r}_\mathrm{f1}=0.886$) are higher than the linear
correlation coefficients of the parameters with $R_\mathrm{z}$
($\overline{r}_\mathrm{0}=0.830$). If a double decay integral
response model is used (with two decay time scales of about 6 and
60 months), the correlations between CME parameters and
$R_\mathrm{z}$ improve ($\overline{r}_\mathrm{f2}=0.906$). The
time delays between CME parameters with respect to Rz are also
well predicted by this model (19/22 = 86\%); the average time
delays are 19 months for the reconstructed and 22 months for the
original time series. The model implies that CMEs are related to
the accumulation of solar magnetic energy. The relationships found
can help to understand the mechanisms at work during the solar
cycle.
\end{abstract}


\section{Introduction}           
     \label{sect:intr}

A coronal mass ejection (CME) is a striking manifestation of solar
activity seen in the solar corona \citep[{\it
e.g.},][]{Gosling93,Webb94}. In a typical CME, billions of tons of
solar magnetized plasma with energy above $10^{32}$ erg can be
pushed into the space \citep[{\it
e.g.},][]{Vourlidas00,Gopalswamy00,Falconer02}. CMEs may cause
strong interplanetary disturbances and geomagnetic storms
\citep{Gonzalez94,Gopalswamy10}. Very high-energy (GeV) particles
generated by CME-driven shocks with large Mach numbers are
hazardous to our modern highly technological equipments
\citep[{\it e.g.},][]{Roussev03,Lee05}.

CMEs originate from large-scale closed magnetic field structures
related to active regions with or without filaments or to
quiescent filaments \citep{Munro79,Chen00,Forbes06,Gopalswamy061}.
The magnetic field in the lower corona is a key element in the
genesis of CMEs \citep[{\it e.g.},][]{Forbes06}. The fastest CMEs
originate because of an instability of strong magnetic fields
($\ge$ 100 G), generally in active regions with sunspots
\citep{Falconer02,Gopalswamy03,Gopalswamy10a} while the high
latitude CMEs are mainly related to quiescent prominence eruptions
\citep{Gopalswamy03}. Sunspots represent one of the most obvious
manifestations of solar magnetic activity \citep{Moradi10} and its
number and characteristics can be considered as a measure of the
energy supplied to the corona \citep{Temmer03}. Studying the
relationships between CME parameters and solar activity,
represented by the international relative sunspot number
($R_\mathrm{z}$), can be useful to understand the mechanisms at
work during the solar activity cycle
\citep{Sakurai76,Antiochos99,Rust03,Wang00,Mittal01,Kilcik11}.

Since the first CME was discovered in data from the Naval Research
Laboratory (NRL) coronagraph mounted on the Orbiting Solar
Obser\-vatory-7 (OSO-7) NASA satellite \citep{Koomen75}, the
properties of CMEs have been carefully examined, particularly
after the advent of the Large Angle and Spectrometric Coronagraph
\citep[LASCO,][]{Brueckner95,Domingo95} on board the Solar and
Heliospheric Observatory (SOHO).

Some CME parameters follow the solar cycle rather well while
others do not \citep[][and references
therein]{Gopalswamy03,Kane06,Ivanov09,Gopalswamy10,Gopalswamy10a,Gerontidou10}.
The number of CMEs per day shows a strong dependence on the solar
cycle
\citep{Hildner76,Cyr00,Gopalswamy03,Gopalswamy061,Gopalswamy10,Cremades07,Gerontidou10}.
The CME speeds have also been found to follow the solar cycle
\citep{Gopalswamy03,Gopalswamy10,Cremades07}. However, the CME
mass and angular width do not exhibit a significant variation with
the solar cycle \citep{Cremades07}. The number of CMEs was shown
to follow well the sunspot number only during the rising phase of
Solar Cycle 23, while there were large fluctuations in the maximum
and the declining phase of this solar cycle
\citep{Gerontidou10,Gopalswamy10}. The CME sources associated with
active regions follow the butterfly diagram and appear at lower
latitudes as the cycle progresses, while those related to
filaments outside active regions tend to migrate towards higher
latitudes \citep[][and references
therein]{Gopalswamy03,Cremades07,Gopalswamy10a}.

It is well known that solar flares  \citep[{\it
e.g.},][]{Aschwanden94,Temmer03}, CMEs
\citep{Gonzalez87,Tsurutani06,Ramesh10,Kilcik11}, and geomagnetic
activities \citep{Wilson90,Echer04,Ramesh10,Kilcik11} often lag
behind sunspot activity ($R_\mathrm{z}$) from several months to a
few years. To have a better understanding of the relationship
between solar ($R_\mathrm{z}$) and geomagnetic activity ($aa$
index), \citet{Du11c} proposed an integral response model to
describe better this relationship. It is found that the
correlation between $aa$ and $R_\mathrm{z}$ yielded by this model
is much higher than that found by a simple point-point
correspondence. Some phenomena can be explained by this model,
such as the significant increase in the $aa$ index over the 20th
century \citep{Feynman78,Cliver98,Demetrescu08,Lukianova09}, the
longer lag times at solar (cycle) maximum than at solar minimum
\citep{Wilson90,Wang00,Echer04}, and the decreasing trend in the
correlation between $aa$ and $R_\mathrm{z}$
\citep{Borello92,Mussino94,Kishcha99,Echer04,Du11b}.

This paper investigates in detail the relationships and time
delays between CME parameters and sunspot numbers
($R_\mathrm{z}$). The CME parameters (speed, acceleration, central
position angle, angular width, mass and kinetic energy) and the
methods used in this paper are described in
Section~\ref{sec:data}. The linear cross-correlations among the
above parameters and their dependencies on $R_\mathrm{z}$ are
first analyzed in Section~\ref{subsec:Correlation}. An integral
response model \citep{Du11c} is then used to study the
relationships and time delays between the CME parameters and
$R_\mathrm{z}$ in Section~\ref{subsec:model1}. In this model, the
output depends not only on the present input, but also on past
values. In a second step, this model is modified and the
relationships between the CME parameters and $R_\mathrm{z}$
improve (see Section \ref{subsec:model2}). Finally, our
conclusions are discussed and summarized in
Section~\ref{sec:Discussions}.  

\section{Data and Methods} \label{sec:data}

The data used in this study comprise monthly-mean sunspot numbers
($R_{\mathrm{z}}$) produced by the Solar Influences Data Analysis
Center (SIDC, \url{http://www.sidc.be/sunspot-data/}) and CME
parameters obtained from the SOHO/LASCO CME catalog identified
since 1996  \citep[\url{http://cdaw.gsfc.\-nasa.\-gov/}, see {\it
e.g.},][]{Gopalswamy00,Yashiro04}. The following CME parameters
are analyzed:
\begin{enumerate}
  \item $V_1$: linear speed obtained by fitting a
straight line to the height-time measurements (in units of km
s$^{-1}$).
  \item $V_{20}$: initial speed obtained by fitting a
  parabola to the height-time measurements.
  \item $V_2$: final speed at the time of final height 
  measurements by fitting a parabola.
  \item $V_{20R}$: speed obtained as above but evaluated
  when the CME is at a height of 20$R_{\odot}$ (solar radii).
  \item $a$: acceleration obtained by fitting a
  parabola to the height-time measurements (in km s$^{-2}$).
  \item $W$: angular width in the plane of the sky (in degrees).
  \item $P_0$: central position angle (CPA, in degrees), from solar north (counterclockwise).
  \item $P_{\mathrm{m}}$: position angle (in
  degrees) at the fastest portion of the leading edge \citep{Yashiro04}.
  \item $M$: mass (in grams).
  \item $E$: kinetic energy (in ergs), obtained from the linear speed ($V_1$) and the CME mass ($M$).
\end{enumerate}

\begin{figure*}[!tb]
 \includegraphics[width=1.8\columnwidth]{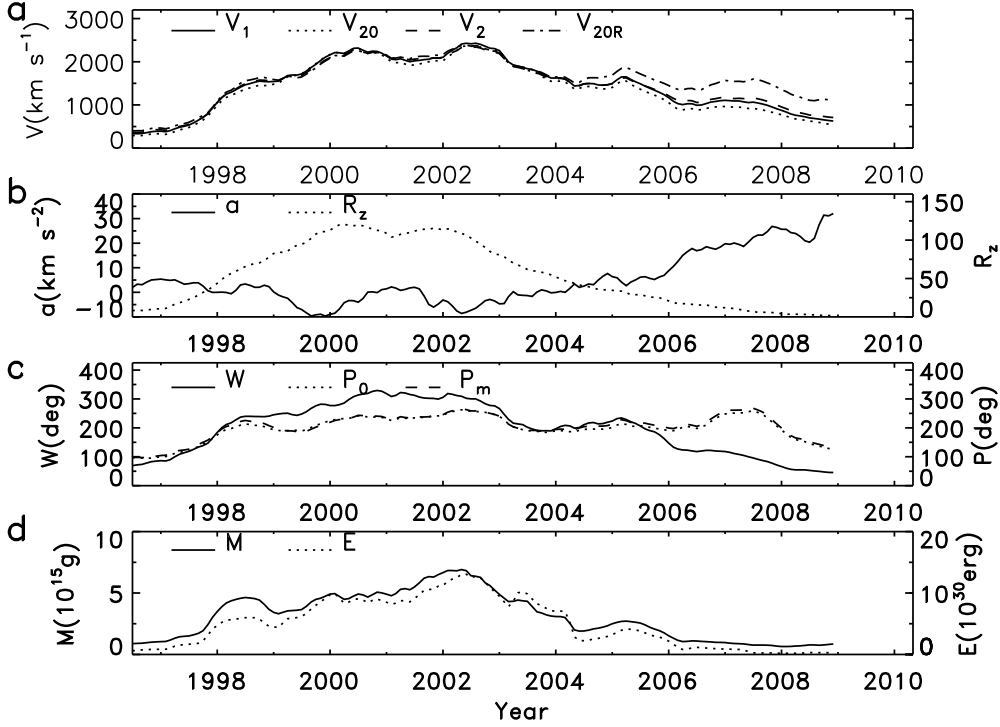}
 \caption{ (a) Smoothed monthly mean $V_1$ (solid line), $V_{20}$ (dotted line), $V_\mathrm{2}$ (dashed line),
     and $V_\mathrm{20R}$ (dash-dotted line) from July 1996 to December 2008.
     (b)  $a$ (solid line) and $R_\mathrm{z}$ (dotted line).
     (c) $W$ (solid line), $P_{0}$ (dotted line), and $P_\mathrm{m}$ (dashed line).
     (d)  $M$ (solid line) and $E$ (dotted line).
     }
 \label{Fig:1}
 \end{figure*}

The above parameters are first integrated over each day and, then,
averaged over each month to obtain the monthly-means of integrated
daily parameters. Such a treatment aims to analyze the total
contribution of CMEs over a month. Ideally, CME projection effects
should be corrected in advance to accurately measure the
parameters \citep{Cremades07,Ivanov09,Gopalswamy10}. As doing this
for all CMEs is a difficult task and there is a small number of
CMEs with $W>120^{\rm o}$ \citep[11\%,][]{Gopalswamy10a},
projection effects have not been corrected in this study. %
In addition, we do not consider the N-S asymmetry of the position
angles ($P_0$ and $P_{\mathrm{m}}$); that is, we use the values of
position angles in terms of co-latitude --- computed from the
solar North pole for the northern hemisphere and from the solar
South pole for the southern hemisphere. The $P_0$ and
$P_{\mathrm{m}}$ values may be greater than $90^{\mathrm{o}}$ due
to the integral contributions of many CMEs in a day. To filter out
high frequency variations in the data, the parameters are smoothed
with the standard 13-month running mean technique (with half
weights at the two ends). The results are shown in
Fig.~\ref{Fig:1} for the period July 1996 to December 2008 (the
time of solar minimum preceding Cycle 24).

\section{Results}
\label{sec:Results}

\subsection{Correlations between CME Parameters} \label{subsec:Correlation}
The relationships among CME parameters and other phenomena have
been well examined in the past. For example, faster CMEs tend to
be wider  \citep[{\it e.g.},][]{Gopalswamy10,Gopalswamy10a}, with
a correlation coefficient of $r=0.44$ between $V_1$ and $W$ for
$V_1>900$ km s$^{-1}$ \citep{Yashiro04}, and wider CMEs generally
have greater mass contents, {$M\propto W^{1.3}$} \cite[{\it
e.g.},][]{Gopalswamy10,Aarnio11}. CMEs with the above-average
speeds {(466 km s$^{-1}$)} decelerate, while those with speeds
well below the average accelerate
\citep{Gopalswamy00,Yashiro04,Gopalswamy061,Gopalswamy10a,Aarnio11}.
Strong CMEs are often accompanied by solar flares, type II radio
bursts, energetic particles, and geomagnetic storms \citep[{\it
e.g.},][]{Gonzalez94,Roussev03,Gopalswamy10}.

 \begin{table*}[!tb]
 \small
 \caption{Correlation coefficients between CME parameters.}
  \label{Tab:tab1}
 \begin{tabular}{llllllllllll}
 \tableline  
 $r$& $V_1$& $V_\mathrm{20}$& $V_\mathrm{2}$&
$V_\mathrm{20R}$&
  $-a$&  $W$& $P_\mathrm{0}$ & $P_\mathrm{m}$ & $M$& $E$ & $R_\mathrm{z}$\\
  \tableline
 $V_1$            &1.    &0.998 &0.998 &0.956 &       0.670 &0.929 &0.771 &0.724  & 0.887 &0.912 & 0.894\\
 $V_\mathrm{20}$  &0.998 &1.    &0.994 &0.942 &       0.696 &0.932 &0.742 &0.690  & 0.893 &0.921 & 0.908\\
 $V_\mathrm{2}$   &0.998 &0.994 &1.    &0.967 &       0.637 &0.923 &0.794 &0.749  & 0.877 &0.899 & 0.888\\
 $V_\mathrm{20R}$ &0.956 &0.942 &0.967 &1.&           0.436 &0.820 &0.886 &0.853  & 0.768 &0.794 & 0.768\\
 $-a$             &0.670 &0.696 &0.637 &0.436 &       1.    &0.805 &0.205 &0.147  & 0.791 &0.793 & 0.795\\
 $W$              &0.929 &0.932 &0.923 &0.820 &       0.805 &1.    &0.630 &0.588  & 0.926 &0.892 & 0.943\\
 $P_\mathrm{0}$   &0.771 &0.742 &0.794 &0.886 &       0.205 &0.630 &1.    &0.994  & 0.569 &0.563 & 0.551\\
 $P_\mathrm{m}$   &0.724 &0.690 &0.749 &0.853 &       0.147 &0.588 &0.994 &1.     & 0.518 &0.502 & 0.485\\
 $M$              &0.887 &0.893 &0.877 &0.768 &       0.791 &0.926 &0.569 &0.518  & 1.    &0.971 & 0.936\\
 $E$              &0.912 &0.921 &0.899 &0.794 &       0.793 &0.892 &0.563 &0.502  & 0.971 &   1. & 0.916\\
 \tableline 
 \end{tabular}
 \end{table*}

In this section, we simply analyze the correlation coefficients
between the CME parameters, as listed in Table~\ref{Tab:tab1}. It
is seen that these values are very high, from 0.77 to 0.99, except
for related to $a$, $P_{\mathrm{0}}$ and $P_{\mathrm{m}}$. The
following can be noted:
\begin{enumerate}
  \item $V_1$ (solid line), $V_{20}$ (dotted line), $V_2$ (dashed line), and $V_{\mathrm{20R}}$ (dash-dotted line)
in Fig.~\ref{Fig:1}a are highly cross-correlated with correlation
coefficients in the range from $r=0.956$ to 0.998. So, if one of
these speeds is well correlated with any other parameter ({\it
e.g.}, $R_{\rm z}$), the others will also be.
  \item $a$ (solid line in Fig.~\ref{Fig:1}b) is inversely correlated
with $V_1$, $V_{20}$, $V_2$, and $V_{\mathrm{20R}}$, $r=-0.436$ to
$-0.696$, implying that CMEs with faster speeds tend to decelerate
faster
\citep{Gopalswamy00,Yashiro04,Gopalswamy061,Gopalswamy10a,Aarnio11}.
  \item $W$ (solid line in Fig.~\ref{Fig:1}c) is well correlated with $V_1$, $V_{20}$, $V_2$,
and $V_{\mathrm{20R}}$, $r=0.820$ to $0.932$, suggesting that CMEs
with faster speeds tend to be wider
\citep{Yashiro04,Gopalswamy10,Gopalswamy10a}.
  \item $W$ is well anti-correlated with $a$, $r=-0.805$, suggesting that wider CMEs tend to
decelerate faster.
  \item Both $P_{\mathrm{m}}$ (dotted line) and $P_0$ (dashed line) in
Fig.~\ref{Fig:1}c are well correlated with $V_1$, $V_{20}$, $V_2$,
and $V_{\mathrm{20R}}$, $r=0.690$ to 0.886, suggesting that faster
CMEs tend to be closer to the solar equator.
  \item $P_{\mathrm{m}}$ is highly correlated with $P_0$,
$r=0.994$, with the former slightly larger than the latter
($5.1^{\mathrm{o}}$).
  \item $W$ is positively correlated with $P_0$ and $P_{\mathrm{m}}$, $r=0.630$ and 0.588, respectively,
suggesting that wider CMEs tend to be closer to the solar equator.
  \item $a$ is almost uncorrelated with both $P_0$ and $P_{\mathrm{m}}$,
$r=-0.205$ and $-0.147$, respectively.
  \item $M$ (solid line in Fig.~\ref{Fig:1}d) is well correlated
with $V_1$, $V_{20}$, $V_2$, and $V_{\mathrm{20R}}$, $r=0.768$ to
$0.893$, suggesting that faster CMEs carry more mass outward.
  \item $M$ is highly correlated with $W$, $r=0.926$, suggesting that
  wider CMEs carry more mass outward
  \citep{Gopalswamy10,Aarnio11}.
  \item $M$ is correlated with $P_0$ and $P_{\mathrm{m}}$, $r=0.569$ and $0.518$, respectively,
suggesting that CMEs closer to the solar equator tend to carry
more mass outward.
  \item $E$ (dotted line in Fig.~\ref{Fig:1}d) is
highly correlated with $M$, $V_1$, $V_{20}$, $V_2$,
 and  $V_{\mathrm{20R}}$ since $E=MV^2/2$ \citep{Vourlidas00}.
  \item $E$ is well correlated with $W$, $r=0.892$, suggesting that wider CMEs carry away more
  energy \citep{Gopalswamy10,Aarnio11}.
  \item $E$ is correlated with $P_0$ and $P_{\mathrm{m}}$, $r=0.563$ and $0.502$, respectively,
suggesting that CMEs closer to the solar equator tend to carry
away more energy.
  \item The correlation coefficients of $P_{\mathrm{m}}$
  are systemically lower than those of $P_0$ with other parameters,
  which may be due to a more non-linear behavior close to the
  leading edge.
  \item The parameters are well correlated with
  $R_{\mathrm{z}}$ (dotted line in Fig.~\ref{Fig:1}b), except for an anti-correlation of $a$ with
  $R_{\mathrm{z}}$.
  \item $a$ is well anti-correlated with $M$ ($r=-0.791$).
\end{enumerate}

In summary, CMEs with faster speeds tend to be wider, to be closer
to the solar equator, to decelerate faster, and to carry more mass
and energy outward. For other properties of CMEs, the readers can
refer to excellent reviews in the literature
\citep{Gopalswamy061,Cremades07,Gopalswamy10,Gerontidou10}.

 \begin{figure}[!tb]
 \includegraphics[width=0.9\columnwidth]{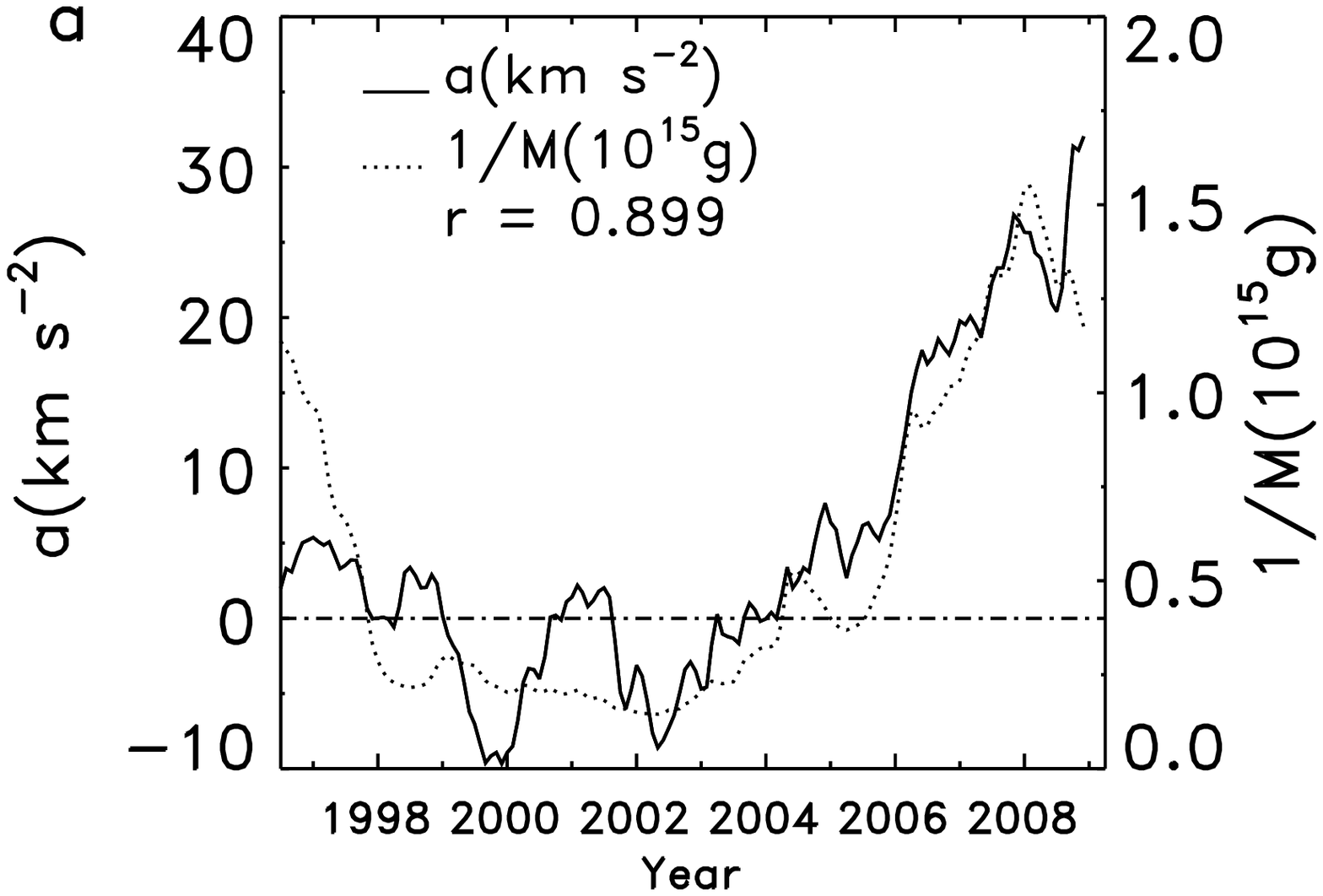}
 \includegraphics[width=0.9\columnwidth]{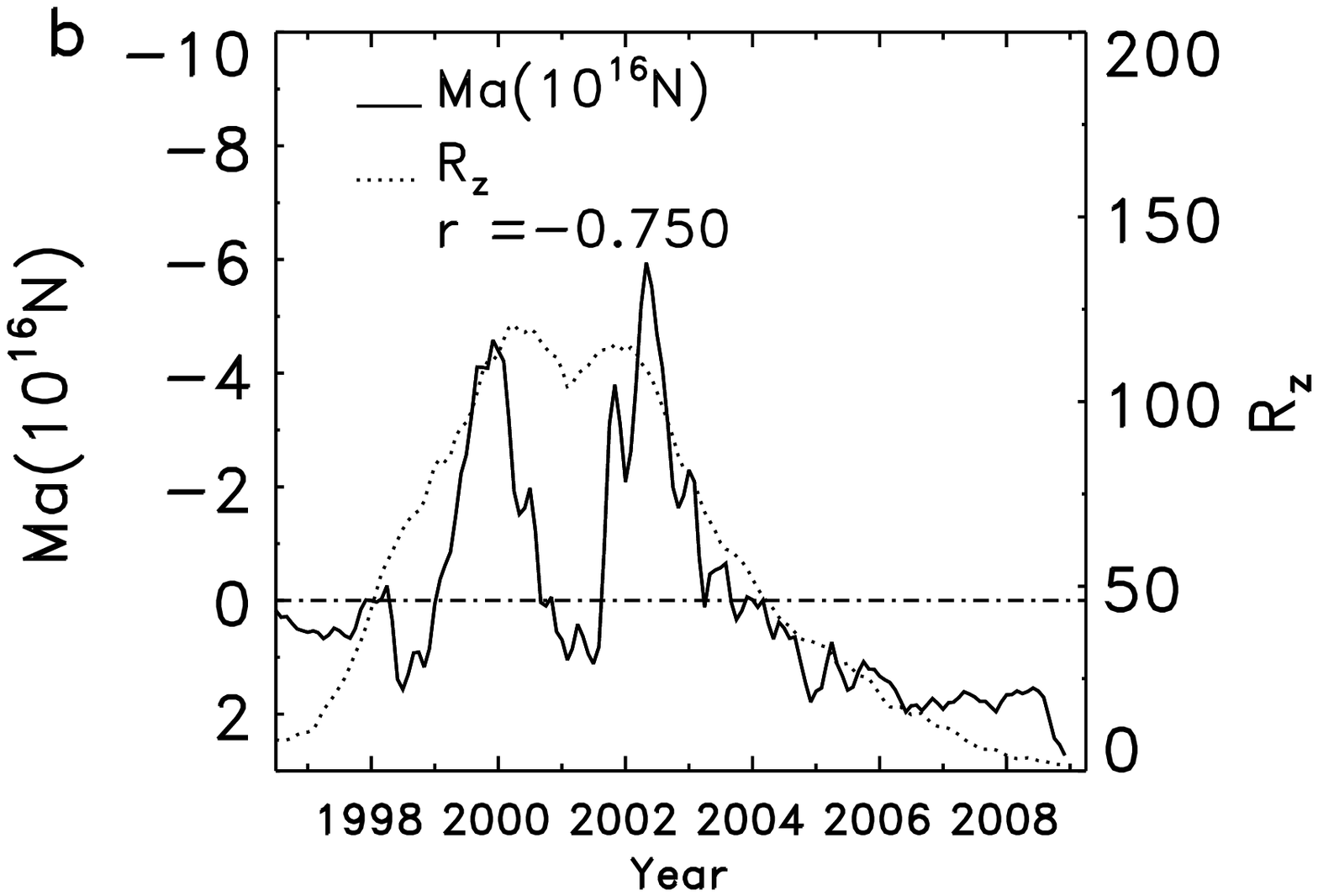}
 \caption{(a) $a$ (solid line) is well correlated with $1/M$ (dotted line), $r=0.899$.
    (b) $Ma$ (solid line) is well anti-correlated with $R_\mathrm{z}$ (dotted line),
    $r=-0.750$.}
 \label{Fig:2}
 \end{figure}

\citet{Aarnio11} pointed out that the mass of a CME is unrelated
to its acceleration. However, it is shown that $a$ is well
anti-correlated with $M$ ($r=-0.791$). In fact, $a$ is highly
correlated with the reciprocal of the mass $1/M$ ($r=0.899$), as
shown in Fig.~\ref{Fig:2}a. It seems to suggest that the CME
motion approximately obeys Newton's second law. Thus, $F=Ma$ can
be understood as the `force' to drive a CME. This force is found
to be well anti-correlated with $R_\mathrm{z}$ ($r=-0.750$), as
shown in Fig.~\ref{Fig:2}b for $Ma$ (solid line) and
$R_\mathrm{z}$ (dotted line). $F>0$ around the solar minima, $F=0$
around $R_\mathrm{z}\sim 41$, while there is not a definite sign
of $F$ around the solar maximum. It is well known that there are
two peaks in $R_\mathrm{z}$ for Cycle 23, the first one (120.8 in
April 2000) being higher than the second one (115.5 in November
2001). One can see in Fig.~\ref{Fig:2}b that there are also two
peaks in $-(Ma)$ for Cycle 23, the first one ($4.6\times 10^{16}$
in December 1999) being lower than the second one ($5.9\times
10^{16}$ in May 2002). Around the two peaks in $R_\mathrm{z}$,
$F<0$, while coincident with the gap between both peaks, $F>0$.

\subsection{Relationships between CME Parameters and $R_\mathrm{z}$
Described by an Integral Response Model} \label{subsec:model1}
\begin{figure*}[!tb]
 \includegraphics[width=1.8\columnwidth]{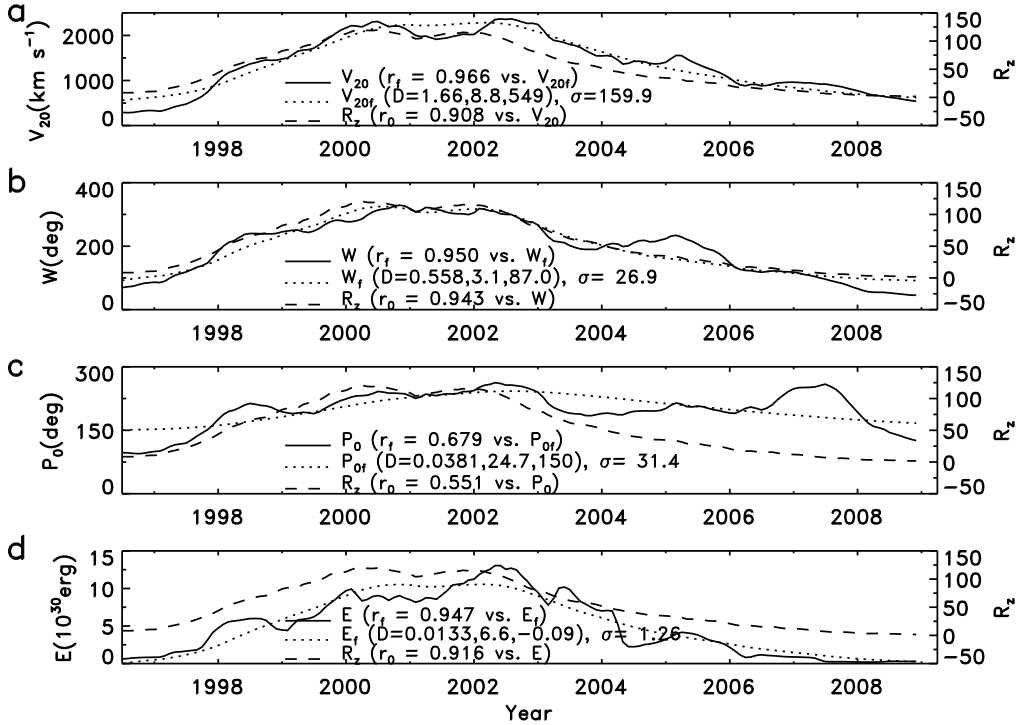}
 \caption{(a)  $V_{20}$ (solid line), $R_\mathrm{z}$ (dashed line), and the reconstructed series
     ($V_\mathrm{20f}$, dotted line)
     by Model I~(Equation~(\ref{Eq:model1})).
     The correlation coefficients of $V_{20}$ with $R_\mathrm{z}$ and $V_\mathrm{20f}$
     are $r_0=0.908$ and $r_\mathrm{f}=0.966$, respectively.
   (b) Similar results for the relationship between $W$ (solid line) and $R_\mathrm{z}$ (dashed line).
   The correlation coefficients of $W$ with $R_\mathrm{z}$ and the reconstructed series
   $W_\mathrm{f}$ (dotted line)  are $r_0=0.943$ and $r_\mathrm{f}=0.950$, respectively.
     (c) Relationship between $P_0$ (solid line) and $R_\mathrm{z}$ (dashed line).
   The correlation coefficients of $P_0$ with $R_\mathrm{z}$ and the reconstructed series
   $P_\mathrm{0f}$ (dotted line)  are $r_0=0.551$ and $r_\mathrm{f}=0.679$, respectively.
     (d) Relationship between $E$ (solid line) and $R_\mathrm{z}$ (dashed line).
   The correlation coefficients of $E$ with $R_\mathrm{z}$ and the reconstructed series
   $E_\mathrm{f}$ (dotted line)  are $r_0=0.916$ and $r_\mathrm{f}=0.947$, respectively.}
 \label{Fig:3}
 \end{figure*}

It is well known that the occurrence of the largest 
CMEs tends to peak some years after the sunspot maximum
\citep{Gonzalez87,Tsurutani06,Ramesh10,Kilcik11}. To explain the
time delays between CME parameters with respect to $R_\mathrm{z}$,
in this section, we employ the following integral response
model~\citep{Du11c}, hereafter Model I, to analyze the
relationships between some typical CME parameters ($V_{20}$, $W$,
$P_0$ and $E$) and $R_\mathrm{z}$,
\begin{eqnarray}
    \label{Eq:model1}
   \begin{array}{lrl}
    y(t)     &=& D\int_{t'=-\infty}^t x(t') e^{-(t-t')/\tau}dt'+y_0,\\
    && {(\rm Model\ I)}\\
   \end{array}
\end{eqnarray}
where $y$ represents one of the CME parameters and
$x=R_\mathrm{z}$, $y_0$ is a constant, $D$ is called the `dynamic
response factor', and $\tau$ is called the `response time scale'
of $y$ to $x$. Figure~\ref{Fig:3}a illustrates the reconstructed
series ($V_\mathrm{20f}$, dotted line) based on this model for
$y=V_{20}$ (solid line) and $x=R_\mathrm{z}$ (dashed line).

It is apparent in Fig.~\ref{Fig:3}a that the reconstructed series
($V_\mathrm{20f}$) reflects well the profile of $V_{20}$. The
correlation coefficient between $V_{20}$ and $V_\mathrm{20f}$
($r_\mathrm{f}=0.966$) is higher than the linear correlation
coefficient between $V_{20}$ and $R_\mathrm{z}$ ($r_0=0.908$). The
fitted parameters of this model are listed in
Table~\ref{Tab:tab2}, in which $\sigma$ refers to the standard
deviation.

 \begin{table*}[!tb]
 \small
 \caption{Fitted parameters of the integral response model (Equation~(\ref{Eq:model1})).}
  \label{Tab:tab2}
 \begin{tabular}{llrrrcc}
 \tableline  
 $y$-$x$& $D$& $\tau$& $y_0$& $\sigma$&
  $r_0$& $r_\mathrm{f}$\\
  \tableline
 $V_{20}$-$R_\mathrm{z}$        & 1.66   & 8.8  & 549    &159.9 &0.908 &0.966\\
 $W$-$R_\mathrm{z}$             & 0.558  & 3.1  & 87.0   & 26.9 &0.943 &0.950 \\
 $P_0$-$R_\mathrm{z}$           & 0.0381 &24.7  & 150    & 31.4 &0.551 &0.679 \\
 $E/10^{30}$-$R_\mathrm{z}$     & 0.0133 & 6.6  &$-0.09$ & 1.26 &0.916 &0.947 \\
 \tableline
 Average                        &        &10.8  &        &      &0.830 &0.886\\
 \tableline 
 \end{tabular}
 \end{table*}

Figure~\ref{Fig:3}b depicts the reconstructed series
($W_\mathrm{f}$, dotted line) based on Model I for $y=W$ (solid
line) and $x=R_\mathrm{z}$ (dashed line). The correlation
coefficient between $W$ and $W_\mathrm{f}$ ($r_\mathrm{f}=0.950$)
is slightly higher than that between $W$ and $R_\mathrm{z}$
($r_0=0.943$). Similarly, the correlation coefficient between
$P_{0}$ and the reconstructed series $P_\mathrm{0f}$ (dotted line)
based on Model I ($r_\mathrm{f}=0.679$) is higher than that
between $P_{0}$ and $R_\mathrm{z}$ ($r_0=0.551$; see
Fig.~\ref{Fig:3}c). The correlation coefficient between $E$ and
the reconstructed series ($E_\mathrm{f}$, dotted line) based on
Model I ($r_\mathrm{f}=0.947$) is higher than that between $E$ and
$R_\mathrm{z}$ ($r_0=0.916$, see Fig.~\ref{Fig:3}d). The above
results are listed in Table~\ref{Tab:tab2}, in which the last row
indicates the relevant averages. The average `response time scale'
is $\overline{\tau}=10.8$ months. The correlation coefficients
increased from 0.830 to 0.886 in average when using Model I.
Therefore, Model I can better describe the relationships between
CME parameters and $R_\mathrm{z}$.

In Model I, $y_0$ is a constant, representing the part of CMEs
that are uncorrelated with $R_\mathrm{z}$ (related to variability
of other solar phenomena), $D$ is the `dynamic response factor' of
$y$ (a CME parameter) to $x (R_\mathrm{z}$), representing the
initial efficiency of CMEs ($\partial y/\partial x|_{t'=t}$), and
$\tau$ is the `response time scale' of $y$ to $x$, representing
that an input (solar activity) may affect the output in the
subsequent time period of about $\tau$ (months) according to an
exponential decay factor ($e^{-(t-t')/\tau}$). It implies that
CMEs are related to the previous accumulation of solar magnetic
energy. As is well known, the upper chromospheric activity indices
tend to lag behind the sunspot number by one to several months,
depending on the index, which was interpreted in terms of active
regions evolving from the photosphere upward
\citep{Bachmann94,Aschwanden94,Temmer03}. Various solar magnetic
activities evolve from the photosphere to upper chromosphere with
different speeds and times. On average, the long-term evolution
times (or periodicities) and the above lag times are reflected in
the `response time scale' ($\overline{\tau}\approx 11$ months).

\subsection{Relationships between CME Parameters and $R_\mathrm{z}$
via a Double Decay Integral Response Model} \label{subsec:model2}
\begin{figure*}[!tb]
 \includegraphics[width=1.8\columnwidth]{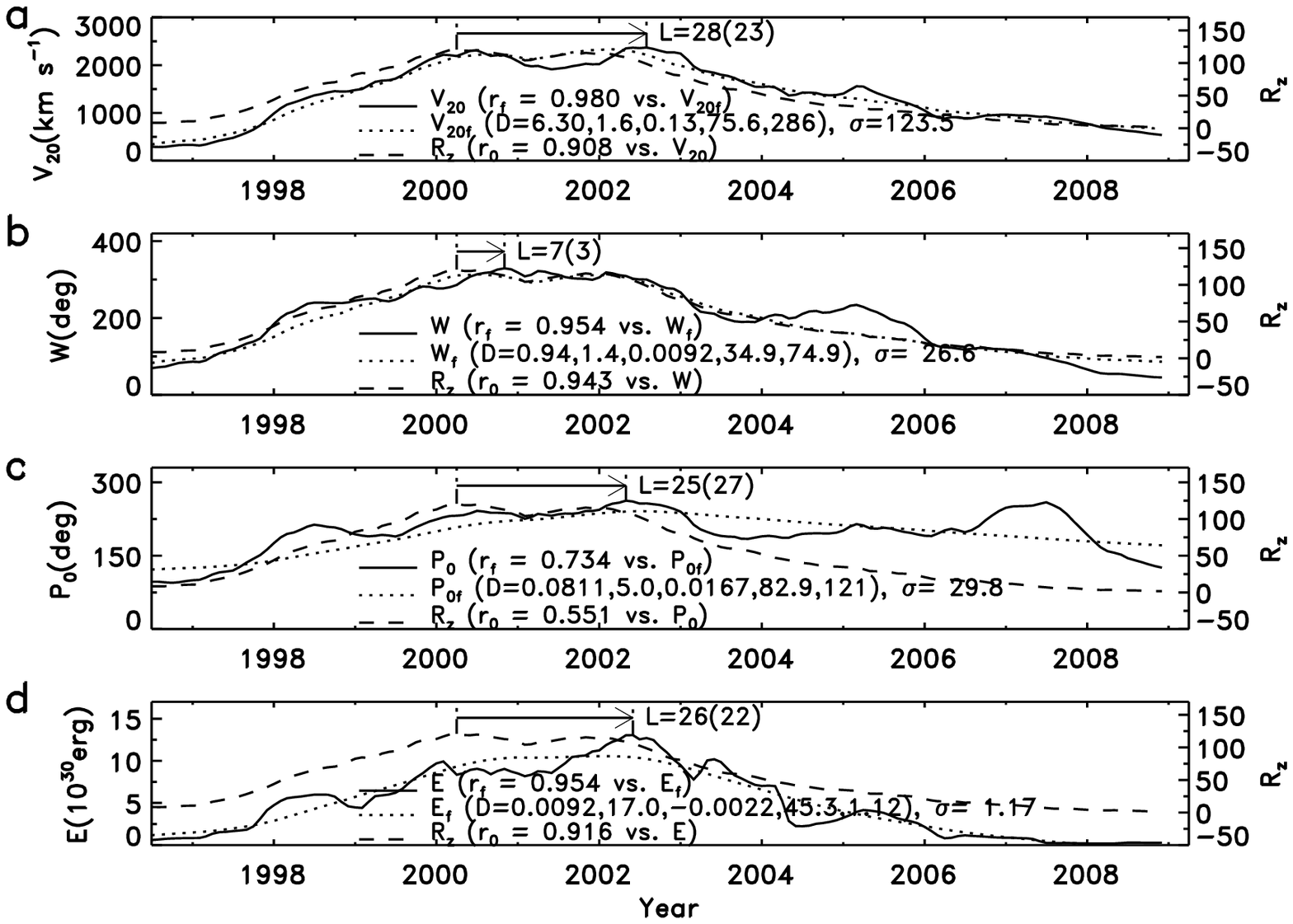}
 \caption{Similar to Fig.~\ref{Fig:3} but using Model II (Equation~(\ref{Eq:model2})).
  (a) The correlation coefficient of $V_{20}$ (solid line)
     with the reconstructed series $V_\mathrm{20f}$ (dotted line) from $R_\mathrm{z}$ (dashed line) by
     Model II
     is now $r_\mathrm{f}=0.980$.
     The lag time of $V_{20}$ ($V_\mathrm{20f}$) with respect to $R_\mathrm{z}$ at the peak of Cycle 23
     is $L_0=28$ ($L_\mathrm{f}=23$ months) months.
  (b) The correlation coefficient of $W$ (solid line)
     with the reconstructed series $W_\mathrm{f}$ (dotted line) from $R_\mathrm{z}$ (dashed line)
     by Model II
     is now $r_\mathrm{f}=0.954$.
     The lag time of $W$ ($W_\mathrm{f}$) with respect to $R_\mathrm{z}$ is $L_0=7$ ($L_\mathrm{f}=3$ months) months.
  (c) The correlation coefficient of $P_{0}$ (solid line)
     with the reconstructed series $P_\mathrm{0f}$ (dotted line) from $R_\mathrm{z}$ (dashed line)
     by Model II
     is now $r_\mathrm{f}=0.734$.
     The lag time of $P_\mathrm{0}$ ($P_\mathrm{0f}$) with respect to $R_\mathrm{z}$ is $L_0=25$
     ($L_\mathrm{f}=27$ months) months.
  (d) The correlation coefficient of $E$ (solid line)
     with the reconstructed series $E_\mathrm{f}$ (dotted line) from $R_\mathrm{z}$ (dashed line)
     by Model II
     is now $r_\mathrm{f}=0.954$.
     The lag time of $E$ ($E_\mathrm{f}$) with respect to $R_\mathrm{z}$ is $L_0=26$ ($L_\mathrm{f}=22$ months) months.
               }
 \label{Fig:4}
 \end{figure*}

As there are often two (or three) peaks in several parameters
associated to solar activity \citep{Gnevyshev67,Feminella97} or
CMEs
\citep{Gonzalez87,Yashiro04,Tsurutani06,Kane06,Cremades07,Ramesh10},
we use the following double decay model (Model II, hereafter),
\begin{eqnarray}
    \label{Eq:model2}
   \begin{array}{lrl}
    y(t)     &=& \int_{t'=-\infty}^t x(t')
    \left[D_1e^{-(t-t')/\tau_1}\right.\\
    &&+\left.D_2e^{-(t-t')/\tau_2}\right]dt'
              +y_0,\\
     && {(\rm Model\ II)}\\
   \end{array}
\end{eqnarray}
to better describe the relationships between the CME parameters
and $R_\mathrm{z}$. The results based on Model II are shown in
Fig.~\ref{Fig:4} and Table~\ref{Tab:tab3}.

One can see in Fig.~\ref{Fig:4}a that the correlation coefficient
of $y=V_{20}$ (solid line) with the reconstructed series
$V_\mathrm{20f}$ (dotted line) from $x=R_\mathrm{z}$ (dashed line)
by Model II is now $r_\mathrm{f}=0.980$, slightly higher than that
of $V_{20}$ with the reconstructed series based on Model I in
Fig.~\ref{Fig:3}a (0.966). The lag time of $V_{20}$ with respect
to $R_\mathrm{z}$ ($L_0=28$ months) is well predicted by this
model ($L_\mathrm{f}=23$ months).

 \begin{table*}[!tb]
 \small
 \caption{Fitted parameters of the double decay integral
response model (Equation~(\ref{Eq:model2})).}
  \label{Tab:tab3}
 \begin{tabular}{llrrrrcccrr}
 \tableline  
 $y$-$x$& $D_1$& $\tau_1$& $D_2$& $\tau_2$& $y_0$& $\sigma$& $r_0$&
$r_\mathrm{f}$
 &$L_0$ &$L_\mathrm{f}$\\
  \tableline
 $V_{20}$-$R_\mathrm{z}$        & 6.30&    1.6&  0.13&      75.6&               286&   123.5&  0.908&  0.980  &28 &23\\
 $W$-$R_\mathrm{z}$             & 0.94&    1.4&  0.0092&    34.9&               74.9&   26.6&  0.943&  0.954  &7  &3\\
 $P_0$-$R_\mathrm{z}$           & 0.0811&  5.0&  0.0167&    82.9&               121&    29.8&  0.551&  0.734  &25 &27\\
 $E/10^{30}$-$R_\mathrm{z}$     & 0.0092&  17.0& $-0.0022$& 45.3&               1.12&   1.17&  0.916&  0.954  &26 &22\\
 \tableline
 Average                        &        &10.8  &        &      &0.830 &0.886\\
 \tableline 
 \end{tabular}
 \end{table*}

Figure~\ref{Fig:4}b illustrates the results for $y=W$ (solid line)
and $x=R_\mathrm{z}$ (dashed line). The correlation coefficient of
$W$ with the reconstructed series $W_\mathrm{f}$ (dotted line) by
Model II is now $r_\mathrm{f}=0.954$, slightly higher than that of
$W$ with the reconstructed series based on Model I in
Fig.~\ref{Fig:3}b ($0.950$). {About half of the} lag time of
$V_{20}$ with respect to $R_\mathrm{z}$ ($L_0=7$ months) is
predicted by this model ($L_\mathrm{f}=3$ months). In
Fig.~\ref{Fig:4}c, the correlation coefficient of $P_{0}$ (solid
line) with the reconstructed series $P_\mathrm{0f}$ (dotted line)
from $R_\mathrm{z}$ (dashed line) by Model II is now
$r_\mathrm{f}=0.734$, slightly higher than that of $P_{0}$ with
the reconstructed series based on Model I in Fig.~\ref{Fig:3}c
(0.679). The lag time of $V_{20}$ with respect to $R_\mathrm{z}$
($L_0=25$ months) is approximately predicted by this model
($L_\mathrm{f}=27$ months). Similarly in Fig.~\ref{Fig:4}d, the
correlation coefficient of $E$ (solid line) with the reconstructed
series $E_\mathrm{f}$ (dotted line) from $R_\mathrm{z}$ (dashed
line) by
Model II 
is now $r_\mathrm{f}=0.954$, slightly higher than that of $E$ with
the reconstructed series based on Model I in Fig.~\ref{Fig:3}d
(0.947). The lag time of $V_{20}$ with respect to $R_\mathrm{z}$
($L_0=26$ months) is approximately predicted by this model
($L_\mathrm{f}=22$ months).

In Table~\ref{Tab:tab3}, the last row indicates the relevant
averages. The average correlation coefficient based on Model II
($\overline{r}_\mathrm{f2}=0.906$) is slightly higher than that
based on Model I ($\overline{r}_\mathrm{f1}=0.886$) and higher
than the linear correlation coefficient
($\overline{r}_\mathrm{0}=0.830$). The two average response time
scales of Model II are about $\overline{\tau}_1=6.3$ and
$\overline{\tau}_2=59.7$ months, one being shorter and the other
one being longer than that for the case of Model I
($\overline{\tau}=10.8$ months). Model I is therefore a simplified
version of Model II.

It has been known that low-energy phenomena ({\it e.g.,} sunspots,
10.7-cm radio flux and Ca K index) are associated with {lower}
atmospheric layers {({\it e.g.,} photosphere, chromosphere and
upper chromosphere, respectively)}, while high energy phenomena
{({\it e.g.,} CMEs) are associated with higher atmospheric layers
({\it e.g.,} corona)} and originate mostly from the solar active
regions \citep{Gosling76,Bachmann94,Feminella97,Sheeley99}. The
solar magnetic structures {(represented by some activity indices)}
may be grouped into two classes: short-lived and long-lived.
Similarly to the case described in the previous section, CMEs are
assumed to be related to the previous accumulation of magnetic
energy, but now with two `response time scales': the first
($\overline{\tau}_1\approx 6$ months) is related to short-lived
weak structures which tend to lag with respect to $R_\mathrm{z}$
by several months \citep{Bachmann94,Temmer03}, and the second
($\overline{\tau}_2\approx 60$ months) is related to long-lived
strong ones which tend to peak later than $R_\mathrm{z}$ by about
a few years \citep{Aschwanden94,Bromund95} and may be related to
the 5.5-year periodicity in solar activity \citep{Das03}.

\section{Discussions and Conclusions}
\label{sec:Discussions}
%
It has been shown that the CME parameters (speed, acceleration,
central position angle, angular width, mass and kinetic energy)
are well correlated, with high correlation coefficients (from 0.77
to 0.99) between most of them. These indicate that CMEs with
faster speeds tend to be wider, to be closer to the solar equator,
to decelerate faster, and to carry more mass and energy outward.
After 2004, $V_\mathrm{20R}$ becomes larger than the other three
speeds ($V_\mathrm{1}$, $V_\mathrm{20}$ and $V_\mathrm{2}$), which
is related to the increasing trend in the acceleration $a$ (solid
line in Fig.~\ref{Fig:1}b). In addition, $a$ is found to be highly
correlated with the reciprocal of mass, $1/M$ ($r=0.899$),
suggesting that the CME motion obeys Newton's second law. The
force to drive a CME is found to be well anti-correlated with
$R_\mathrm{z}$ ($r=-0.750$).

The relationships between some typical CME parameters ($V_{20}$,
$W$, $P_\mathrm{0}$ and $E$) and sunspot numbers ($R_\mathrm{z}$)
can be well described by an integral response model
(Equation~(\ref{Eq:model1})) --- Model I. The correlation
coefficients between the CME parameters and $R_\mathrm{z}$
increase by 6.7\% from $\overline{r}_\mathrm{0}=0.830$ to
$\overline{r}_\mathrm{f1}=0.886$ on average when using this model
(Table~\ref{Tab:tab2}). In this model, the output $y(t)$ depends
not only on the present input $x(t)$ but also on its past values
according to an exponential decay factor $ e^{-(t-t')/\tau}$. The
earlier the input, the less it contributes to the output. This
implies that CMEs are related to the previous accumulation of
solar magnetic energy with a mean `response time scale' of about
$\overline{\tau}\approx 11$ months. Apart from short-term
variations, parameters related to solar magnetic activity have
also long-term components evolving from the photosphere to the
upper chromosphere. The magnetic energy (in some events) can be
stored to be released later instead of being released
instantaneously. As the CME parameters are well cross-correlated,
{similar conclusions can also be drawn when using the other six
CME parameters in Table \ref{Tab:tab1}.}

Using the double decay integral response model
(Equation~(\ref{Eq:model2})) --- Model II, the above correlations
improve ($\overline{r}_\mathrm{f2}=0.906$). {These results mean
that about $\overline{r}_0^2=68.9\%$ (coefficient of
determination) of the variations in the CME parameters can be
explained by the variation in $R_\mathrm{z}$ through a linear
relationship. While about $\overline{r}^2_\mathrm{f1}=78.5\%$
($\overline{r}^2_\mathrm{f2}=82.1\%$) of the variations in the CME
parameters can be explained by the variation in $R_\mathrm{z}$
through Model I (II).} Besides, the time delays of CME parameters
with respect to $R_\mathrm{z}$ ($\overline{L}_0=22$ months) can
also be well predicted by the model ($\overline{L}_\mathrm{f}=19$
months), {$\overline{L}_\mathrm{f}/\overline{L}_0=86\%$}.
Therefore, Model I or II can better describe the relationships
between CME parameters and $R_\mathrm{z}$.

Model II implies that the relationships between CME parameters and
sunspot (magnetic field) activity depend on two exponential decay
factors, one with a shorter time scale ($\overline{\tau}_1=6.3$
months) and another one with a longer time scale
($\overline{\tau}_2=59.7$ months). This model is related to the
two (or three) peaks often present in solar activity indicators
\citep[$R_\mathrm{z}$, flares, radio, and X-ray
fluxes,][]{Gnevyshev67,Feminella97}, CMEs
\citep{Gonzalez87,Tsurutani06,Ramesh10,Kilcik11}, geomagnetic
activities or CME-related storms \citep{Gonzalez87,Tsurutani06},
one near the peak in $R_\mathrm{z}$ and another one a few years
later. A double peak in these activity indicators suggests that
there are two sources or two decay time scales.

It has long been recognized that low-energy phenomena (slow CMEs)
are associated with deeper atmospheric layers (erupting
prominences) and tend to follow the sunspot activity, while high
energy phenomena (fast CMEs) originate mostly from active regions
and tend to lag behind the sunspot activity
\citep{Gosling76,Feminella97,Sheeley99}. The two `response time
scales' in Model II ($\overline{\tau}_1\approx 6$ months and
$\overline{\tau}_2\approx 60$ months) may be related to the
short-lived weak magnetic structures {(low energy activity
phenomena)} which tend to lag behind $R_\mathrm{z}$ by several
months \citep{Bachmann94,Temmer03} and the long-lived strong ones
which tend to peak later than $R_\mathrm{z}$ by about 2-3 years
\citep{Aschwanden94,Bromund95}. This reflects the fact that solar
magnetic activity evolves from the photosphere to the upper
chromosphere with different speeds and different time scales (or
periodicities). The `response time scale' ($\overline{\tau}\approx
11$ months) in Model I is the mean effect of the two
($\overline{\tau}_1\approx 6$ months and $\overline{\tau}_2\approx
60$ months) in Model II.

Solar structures near the solar equator are more active than those
near the solar poles, which may be related to the faster solar
rotation at lower latitudes. According to the above model(s), the
stronger the inputs ($x$), the more they contribute to the outputs
($y$), and the longer the lag time of $y$ with respect to $x$.
Large active regions of complex magnetic structure are long-lived
and produce intense CMEs which have long time delays with respect
to $R_\mathrm{z}$. Therefore, active regions close to the solar
equator tend to generate CMEs with faster speeds which carry more
energy outward.

As pointed out in Section~\ref{sec:data}, the above results may be
affected by CME projection effects, because the CME parameters are
measured in the plane of the sky. More accurate results could be
obtained using data from the Solar Terrestrial Relations
Observatory (STEREO) as we can deduce the three-dimensional nature
of CMEs. As full halo CMEs account only for 3.6\% of all CMEs and
CMEs with $W>120^{\rm o}$ account for 11\% of all CMEs
\citep{Gopalswamy10a}, and in the present study we use the
integrated values of CME parameters, we expect that our general
conclusions are not significantly influenced by projection
effects.

The main points of this study can be summarized as follows,
\begin{enumerate}
  \item The acceleration of a CME ($a$) is highly correlated with the
reciprocal of its mass ($M$), $r=0.899$.
 \item The force ($Ma$) to drive a CME is well anti-correlated
 with the sunspot number ($R_\mathrm{z}$), $r=-0.750$.
  \item The relationships between CMEs and $R_\mathrm{z}$
can be well described by an integral response model
(Equation~(\ref{Eq:model1})) with a decay time scale of about 11
months. The correlation coefficients of CME parameters with the
reconstructed series based on this model
($\overline{r}_\mathrm{f1}=0.886$) are higher than those based
only on $R_\mathrm{z}$ ($\overline{r}_\mathrm{0}=0.830$).
  \item A double decay integral response model
(Equation~(\ref{Eq:model2})) with two decay time scales of about 6
and 60 months shows better correlations between CME parameters and
$R_\mathrm{z}$ ($\overline{r}_\mathrm{f2}=0.906$).
  \item The time delays of CME parameters with respect to $R_\mathrm{z}$
  can also be well predicted by the model {(86\%)}. 
\end{enumerate}

\section*{Acknowledgments}
The author is grateful to an anonymous referee and the editor for
suggestive and helpful comments which improved the original
version of the manuscript. This work is supported by National
Natural Science Foundation of China (NSFC) through grant Nos
10973020, 40890161 and 10921303, and National Basic Research
Program of
China through grant No. 2011CB811406. 
The International sunspot number is produced by SIDC, RWC Belgium,
World Data Center for the Sunspot Index, Royal Observatory of
Belgium. The CME catalog is generated and maintained at the CDAW
Data Center by NASA and the Catholic University of America in
cooperation with the Naval Research Laboratory. SOHO is a project
of international cooperation between ESA and NASA.



\begin{thebibliography}{}
\bibitem[Aarnio et al.(2011)]{Aarnio11}  Aarnio, A. N., Stassun, K. G., Hughes, W. J., \& McGregor, S. L.
  2011, \solphys,  268, 195
\bibitem[Antiochos et al.(1999)]{Antiochos99}  Antiochos, S. K., DeVore, C. R., \& Klimchuk, J. A.
  1999, \apj,  268, 485
\bibitem[Aschwanden(1994)]{Aschwanden94}  Aschwanden, M. J.
  1994, \solphys,  152, 53
\bibitem[Bachmann \& White(1994)]{Bachmann94}  Bachmann,  K. T., \& White, O. R.
  1994, \solphys,  150, 347
\bibitem[Borello-Filisetti et al.(1992)]{Borello92}
  Borello-Filisetti, O., Mussino, V., Parisi, M., \& Storini, M.
  1992, Ann. Geophys.,  10, 668
\bibitem[Bromund et al.(1995)]{Bromund95}
  Bromund, K. R., McTiernan, J. M., \& Kane, S. R.
  1995, \apj,  455, 733
\bibitem[Brueckner et al.(1995)]{Brueckner95}
    Brueckner, G. E., Howard, R. A., Koomen, M. J.,
    et al.
  1995, \solphys,  162, 357
\bibitem[Chen \& Shibata(2000)]{Chen00}
  Chen, P. F., \& Shibata, K.
  2000, \apj,  545, 524
\bibitem[Clilverd et al.(1998)]{Cliver98}
  Cliver, E.W., Boriakoff, V., \& Feynman, J.
  1998, \grl,  25, 1035
\bibitem[Cremades \& St. Cyr(2007)]{Cremades07}
  Cremades, H., \& St. Cyr, O. C.
  2007, Adv. Space Res.,  40, 1042
\bibitem[Das \& Nag(2003)]{Das03}
  Das, T. K., \& Nag, T. K.
  2003, Bull. Astr. soc. India,  31, 1
\bibitem[Demetrescu \& Dobrica(2008)]{Demetrescu08}
  Demetrescu, C., \& Dobrica, V.
  2008, \jgr,  113, A02103
\bibitem[Domingo et al.(1995)]{Domingo95}
  Domingo, V., Fleck, B., \& Poland, A. I.
  1995, \solphys,  162, 1
\bibitem[Du(2011a)]{Du11c}
  Du, Z. L.
  2011a, Ann. Geophys.,  29, 1005
\bibitem[Du(2011b)]{Du11b}
  Du, Z. L.
  2011b, Ann. Geophys.,  29, 1341
\bibitem[Echer et al.(2004)]{Echer04}
  Echer, E., Gonzalez, W. D., Gonzalez, A. L. C., Prestes, A., Vieira, L. E. A., dal Lago, A.
      Guarnieri, F. L., \& Schuch, N. J.
  2004, J. Atmos. Solar Terr. Phys.,  66, 1019
\bibitem[Falconer et al.(2002)]{Falconer02}
  Falconer, D. A., Moore, R. L., \& Gary, G. A.
  2002, \apj,  569, 1016
\bibitem[Feminella \& Storini(1997)]{Feminella97}
  Feminella, F., \& Storini, M.
  1997, \aap,  322, 311
\bibitem[Feynman \& Crooker(1978)]{Feynman78}
  Feynman, J., \& Crooker, N. U.
  1978, \nat,  275, 626
\bibitem[Forbes et al.(2006)]{Forbes06}
  Forbes, T. G., Linker, J. A., Chen, J.,
     et al.
  2006, \ssr,  123, 251
\bibitem[Gerontidou et al.(2010)]{Gerontidou10}
  Gerontidou, M., Mavromichalaki, H., Asvestari, E., Belov, A., \& Kurt, V.
  2010, AIP Conf. Proc.,  1203, 115
\bibitem[Gnevyshev(1967)]{Gnevyshev67}
  Gnevyshev, M. N.
  1967, \solphys,  1, 107
\bibitem[Gonzalez \& Tsurutani(1987)]{Gonzalez87}
  Gonzalez, W. D., \& Tsurutani, B. T.
  1987, Planet. Space Sci.,  35, 1101
\bibitem[Gonzalez et al.(1994)]{Gonzalez94}
  Gonzalez, W. D., Joselyn, J. A., Kamide, Y.,
     et al.
  1994, \jgr,  99, 5771
\bibitem[Gopalswamy(2006)]{Gopalswamy061}
  Gopalswamy, N.
  2006, J. Astrophys. Astr.,  27, 243
\bibitem[Gopalswamy(2010)]{Gopalswamy10}
  Gopalswamy, N.
  2010, In Coronal Mass Ejections: a Summary of
Recent Results, Proc. 20th National Solar Physics Meeting,
Papradno, Slovakia, 108.
\bibitem[Gopalswamy et al.(2010)]{Gopalswamy10a}
   Gopalswamy, N., Akiyama, S., Yashiro, S., M\"akel\"a, P.:
   2010. In: Hasan, S.S., Rutten, R.J. (eds.) Magnetic Coupling between the
Interior and Atmosphere of the Sun; Astrophys. and Space Scien.
Proc., 289.
\bibitem[Gopalswamy et al.(2000)]{Gopalswamy00}
    Gopalswamy, N., Lara, A., Lepping, R. P.,
  2000, \grl,  27, 145
\bibitem[Gopalswamy et al.(2003)]{Gopalswamy03}
   Gopalswamy, N., Lara, A., Yashiro, S., Nunes, S., Howard, R.A.:
  2003. In: Wilson, A. (ed.) Solar variability as an input to the
  Earth's environment, SP-535, ESA, Noordwijk, 403.
\bibitem[Gosling(1993)]{Gosling93}
  Gosling, J. T.
  1993, \jgr,  98, 18937
\bibitem[Gosling et al.(1976)]{Gosling76}
  Gosling, J. T., Hildner, E., MacQueen, R. M.,
  et al.
  1976, \solphys,  48, 389
\bibitem[Hildner et al.(1976)]{Hildner76}
  Hildner, E., Gosling, J. T., MacQueen, R. M.,
  et al.
  1976, \solphys,  48, 127
\bibitem[Ivanov et al.(2009)]{Ivanov09}
  Ivanov, E. V., Rudenko, G. V., \& Fainshtein, V. G.
  2009, Geomag.  Aeron.,  49, 1096
\bibitem[Kane(2006)]{Kane06}
  Kane, R. P.
  2006, \solphys,  233, 107
\bibitem[Kishcha et al.(1999)]{Kishcha99}
  Kishcha, P. V., Dmitrieva, I. V., \& Obridko, V. N.
  1999, J. Atmos. Solar Terr. Phys.,  61, 799
\bibitem[Kilcik et al.(2011)]{Kilcik11}
  Kilcik, A., Yurchyshyn, V. B., Abramenko, V.,
     et al.
  2011, \apj,  727, 44
\bibitem[Koomen et al.(1975)]{Koomen75}
  Koomen, M. J., Detwiler, C. R., Brueckner, G. E.,
     et al.
  1975, Appl. Opt.,  14, 743
\bibitem[Lee(2005)]{Lee05}
  Lee, M. A.
  2005, Astrophys. J. supl.,  158, 38
\bibitem[Lukianova et al.(2009)]{Lukianova09}
  Lukianova, R.,  Alekseev, G.,  Mursula, K.
  2009, \jgr,  114, A02105
\bibitem[Mittal \& Narain(2010)]{Mittal01}
   Mittal, N., \& Narain, U.
  2010, J. Atmos. Solar Terr. Phys.,  72, 643
\bibitem[Moradi et al.(2010)]{Moradi10}
   Moradi, H., Baldner, C., Birch, A. C.,
    et al.
  2010, \solphys,  267, 1
\bibitem[Munro et al.(1979)]{Munro79}
   Munro, R. H., Gosling, J. T., Hildner, E.,
    et al.
  1979, \solphys,  61, 201
\bibitem[Mussino et al.(1994)]{Mussino94}
   Mussino, V., Borello-Filisetti, O.,  Storini, M., \& Nevanlinna, H.
  1994, Ann. Geophys.,  12, 1065
\bibitem[Ramesh(2010)]{Ramesh10}
   Ramesh, K. B.
  2010, \apj,  712, L77
\bibitem[Roussev et al.(2003)]{Roussev03}
   Roussev, I. I., Gombosi, T. I., Sokolov, I. V.,
    et al.
  2003, \apj,  595, L57
\bibitem[Rust(2003)]{Rust03}
   Rust, D. M.
  2003, Adv. Space Res.,  32, 1895
\bibitem[Sakurai(1976)]{Sakurai76}
   Sakurai, T.
  1976, \pasj,  28, 177
\bibitem[Sheeley et al.(1999)]{Sheeley99}
   Sheeley, N. R., Walters, J. H., Wang, Y. -M.,  Howard, R. A.
  1999, \jgr,  104, 24739
\bibitem[St. Cyr et al.(2000)]{Cyr00}
   St. Cyr, O. C., Plunkett, S. P., Michels, D. J.,
     et al.
  2000, \jgr,  105, 18169
\bibitem[Temmer et al.(2003)]{Temmer03}
   Temmer, M., Veronig, A., Hanslmeier, A.
  2003, \solphys,  215, 111
\bibitem[Tsurutani et al.(2006)]{Tsurutani06}
   Tsurutani, B. T.,  Gonzalez, W. D., Gonzalez, A. L. C.,
      et al.
  2006, \jgr,  111, A07S01
\bibitem[Vourlidas et al.(2000)]{Vourlidas00}
   Vourlidas, A., Subramanian, P., Dere, K. P., Howard, R. A.
  2000, \apj,  534, 456
\bibitem[Wang et al.(2000)]{Wang00}
   Wang, Y.-M., Lean, J., \& Sheeley, N.R.
  2000, \grl,  27, 505
\bibitem[Webb \& Howard(1994)]{Webb94}
   Webb, D. F., \& Howard, R. A.
  1994, \jgr,  99, 4201
\bibitem[Wilson(1990)]{Wilson90}
   Wilson, R. M.
  1990, \solphys,  125, 143
\bibitem[Yashiro et al.(2004)]{Yashiro04}
   Yashiro, S., Gopalswamy, N., Michalek, G.,
   et al.
  2004, \jgr,  109, A07105
\end{thebibliography}
\end{document}